\documentclass[apjl]{emulateapj}

\usepackage{graphics}
\usepackage{apjfonts}
\usepackage{mathptmx}
\usepackage{natbib}

\newcommand{\hi}{H\,{\sc i}\rm}
\newcommand{\hii}{H\,{\sc ii}\rm}
\newcommand{\hei}{He\,{\sc i}\rm}
\newcommand{\heii}{He\,{\sc ii}\rm}
\newcommand{\siii}{[S\,{\sc iii}]}
\newcommand{\nii}{[N\,{\sc ii}]}
\newcommand{\none}{[N\,{\sc i}]}
\newcommand{\oiii}{[O\,{\sc iii}]}
\newcommand{\oii}{[O\,{\sc ii}]}
\newcommand{\oi}{[O\,{\sc i}]}
\newcommand{\sii}{[S\,{\sc ii}]}
\newcommand{\ariii}{[Ar\,{\sc iii}]}
\newcommand{\neiii}{[Ne\,{\sc iii}]}
\newcommand{\feii}{Fe\,{\sc ii}\rm}
\newcommand{\feiii}{[Fe\,{\sc iii}]}
\newcommand{\ariv}{[Ar\,{\sc iv}]}
\newcommand{\cliii}{[Cl\,{\sc iii}]}
\newcommand{\cii}{C\,{\sc ii}\rm}
\newcommand{\silii}{Si\,{\sc ii}\rm}

\newcommand{\te}{$T_e$}

\newcommand{\lin}{$\,\lambda$}
\newcommand{\llin}{$\,\lambda\lambda$}

\newcommand{\oh}{12\,+\,log(O/H)\,=\,}
\newcommand{\ch}{12\,+\,log(C/H)\,=\,}

\slugcomment{}

\shorttitle{Inner \hii\/ regions of M101}
\shortauthors{Bresolin}

\begin{document}


\title{The oxygen abundance in the inner H\,{\small II} regions of M101.\\ Implications for the calibration of strong-line metallicity indicators$^1$\\[2mm]}\footnotetext[1]{The data presented herein were obtained at the W.M. Keck Observatory, which is operated as a scientific partnership among the California Institute of Technology, the University of California and the National Aeronautics and Space Administration. The Observatory was made possible by the generous financial support of the W.M. Keck Foundation.}

\author{Fabio Bresolin} \affil{Institute for Astronomy, 2680 Woodlawn
Drive, Honolulu, HI 96822; bresolin@ifa.hawaii.edu}

\begin{abstract}
I present deep spectroscopy of four \hii\/ regions in the inner, metal-rich zone of the spiral galaxy M101 obtained with the LRIS spectrograph at the Keck telescope. From the analysis of the collisionally excited lines in two of the target \hii\/ regions, H1013 and H493,
I have obtained oxygen abundances \oh8.52 and \oh8.74, respectively. These measurements extend the determination of the oxygen abundance gradient of M101 via the direct method to only 3\,kpc from the center. The intensity of the \cii\lin4267 line 
in H1013 leads to a carbon abundance \ch8.66, corresponding to nearly twice the solar value. From a comparison of the continuum temperature derived from the Balmer discontinuity, $T(Bac)\,=\,5000$~K, and the line temperature derived from \oiii\,\lin4363/\lin5007, $T$\oiii\,=\,7700~K, an average temperature $T_0$\,=\,5500~K and a mean square temperature fluctuation $t^2$\,=\,0.06 have been derived. Accounting for the spatial inhomogeneity in temperature
raises the oxygen abundance obtained from the oxygen auroral lines to \oh8.93.
These findings are discussed in the context of the calibration of strong-line metallicity indicators, in particular of the upper branch of $R_{23}$. There is no evidence for the strong abundance biases arising from temperature gradients predicted theoretically for metal-rich \hii\/ regions.

\end{abstract}

 

\section{Introduction}

As probes of the gas-phase chemical composition of star-forming galaxies,
\hii\/ regions are crucial targets in the quest to understand the chemical
evolution of the Universe. Through the analysis of bright \hii\/ regions in external spiral and irregular galaxies it is, in principle, rather straightforward to measure the abundances of elements such as oxygen, nitrogen, neon, argon and sulphur, and therefore constrain models of nucleosynthesis in massive stars and of galactic chemical evolution.

In recent years a great deal of interest has been raised concerning the derivation 
of chemical abundances in metal-rich \hii\/ regions, i.e.~those where the 
oxygen abundance reaches or exceeds that measured in the Sun [$\rm 12+log(O/H)_\odot=8.66$, \citealt{Asplund:2004}]\footnote{The terms {\em metallicity} and {\em oxygen abundance} will be used interchangeably throughout the paper. Abundances are measured relative to hydrogen.}. This work is motivated by the importance 
of obtaining accurate abundances for the measurement and the interpretation of galactic chemical abundance gradients (\citealt{Pilyugin:2004,Chiappini:2003,Carigi:2005}), and for constraining the metal-rich end of the luminosity-metallicity 
(\citealt{Kobulnicky:2004,Salzer:2005,Maier:2005,Lamareille:2006}) and mass-metallicity (\citealt{Tremonti:2004,Savaglio:2005,Erb:2006}) relations of galaxies up to redshifts $z \sim 2$. A large, solar-like oxygen content is inferred for 
star-forming galaxies both locally (\citealt{Bresolin:2004,Pilyugin:2006}) and at high redshift (\citealt{Shapley:2004}).
In addition, empirical metallicity trends in the Wolf-Rayet WC/WN ratio 
(\citealt{Meynet:2005, Massey:2003}) and in the Cepheid Period-Luminosity Relation (\citealt{Sakai:2004}) depend critically on the reliability of 
the abundances of metal-rich \hii\/ regions.

The difficulty of measuring chemical abundances from the classic analysis of optical forbidden lines (collisionally excited) in high-metallicity \hii\/ regions derives from the increased proportion of gas coolants, in particular oxygen, which emits line radiation mostly via the fine-structure \oiii\llin52,\,88\,$\mu$m lines. As a consequence of 
the reduced gas temperature, the auroral-to-nebular line ratios [e.g.~\oiii\lin4363/(\lin4959\,+\,\lin5007)], employed to determine the  electron temperature $T_e$, become progressively smaller with increasing metallicity, due to the exponential dependence of the collisionally excited line emissivity on $T_e$. Generally, at metallicities approaching solar the \oiii\lin4363 auroral line is too faint to be observed in extragalactic nebulae even with 10m-class telescopes, as a result of the 
strong \te\/ dependence, and of the fact that the far-IR \oiii\/ lines become stronger at the expense of the optical lines.
However,  auroral lines from other ionic species can be detected in the optical spectra of \hii\/ regions at relatively high abundances, such as \siii\lin6312 and \nii\lin5755, as a consequence of their larger Boltzmann factors $exp(-\chi/kT_e)$
($\chi$ is the energy difference, a few eV, between the two levels involved in an auroral line transition). Still, the detection of these diagnostic lines in extragalactic nebulae requires high surface brightness and, generally, 10m-class telescope capabilities.

With the aid of photoionization models (\citealt{Stasinska:1978,Garnett:1992}), one can relate the  electron temperature derived for the \siii\/ or \nii\/ emitting regions [from \siii\lin6312/(\lin9069\,+\,\lin9532) and \nii\lin5755/(\lin6548\,+\,\lin6583), respectively] to the temperature of the \oiii\/ emitting region. When \lin4363 is unavailable, this allows the measurement of the oxygen abundance from the intensity of the strongest lines only (\llin4959,\,5007), having fixed the line emissivity from the knowledge of the electron temperature. This technique has been recently applied by \citet{Bresolin:2004, Bresolin:2005} in Keck and VLT studies of the chemical abundances of samples of extragalactic metal-rich \hii\/ regions. 

The availability of direct abundances is essential for the empirical calibration of the so-called {\em strong-line methods}, in which the strength of easily observed nebular emission lines  are related to the metallicity of the emitting regions. Among the various methods proposed, I recall $R_{23}$\,=\,(\oii\lin3727 + \oiii\llin4959,\,5007)/H$\beta$ (\citealt{Pagel:1979}), N2\,=\,\nii\lin6583/H$\alpha$ (\citealt{Storchi-Bergmann:1994,Denicolo:2002}), O3N2\,=\,(\oiii\lin5007/H$\beta$)/(\nii\lin6583/H$\alpha$) (\citealt{Alloin:1979,Pettini:2004}), $S_{23}$\,=\,(\sii\llin6717,\,6731 + \siii\llin9069,\,9532)/H$\beta$ (\citealt{Vilchez:1996,Diaz:2000}), the P-method (\citealt{Pilyugin:2000,Pilyugin:2005}) and Ar$_3$O$_3$\,=\,\ariii\lin7135/\oiii\lin5007, S$_3$O$_3$\,=\,\siii\lin9069/\oiii\lin5007 (\citealt{Stasinska:2006}). 
For in-depth discussions of these abundance indicators the reader is referred to these papers, as well as to \citet{Kewley:2002}, \citet{Perez-Montero:2005} and \citet{Dopita:2006a}.
The strong-line methods allow the estimation of nebular chemical abundances in cases where a direct approach, based on the detection of the auroral lines, is not feasible, and in general when the signal-to-noise ratio of the spectra is sufficient to measure only a restricted number of emission lines. This approach is widely adopted for the determination of abundance gradients in galaxies (\citealt{Pilyugin:2004, Pilyugin:2006}) and of oxygen abundances in intermediate- and high-redshift star-forming galaxies (\citealt{Shapley:2004,Savaglio:2005,Mouhcine:2006}). The popularity of these metallicity indicators is also due to the increasing necessity of estimating chemical abundances for galaxies observed as part of large spectroscopic surveys, such as the Sloan Digital Sky Survey (\citealt{Tremonti:2004,Liang:2006,Shi:2006}).

Calibrations of strong-line methods based on grids of photoionization models (e.g.~\citealt{Edmunds:1984, McGaugh:1991, Zaritsky:1994, Charlot:2001,Kewley:2002}) provide abundances that are, at the metal-rich end, considerably larger (up to 0.5 dex) than those obtained from direct measurements (\citealt{Castellanos:2002,Kennicutt:2003,Garnett:2004a,Bresolin:2004, Bresolin:2005}) or from \te-calibrated empirical methods.  Mixing \te-based abundances (routinely done at low metallicity) with abundances from model grids (at large metallicity) leads then to spurious discontinuities in abundance diagnostic diagrams or breaks in the luminosity-metallicity relation (\citealt{Nagao:2006,Lee:2006}).
The reasons for the discrepancy are still currently poorly known: additional sources of heating need to be postulated  in order to reproduce theoretically the measured \oiii\lin4363 line strength (\citealt{stasinska:1999,luridiana:1999}).

There are a number of open issues regarding the study of metal-rich \hii\/ regions that motivate further investigation of ionized  nebulae in nearby galaxies. First of all is the assessment of the importance of the effects of temperature 
gradients in \hii\/ regions and the resulting abundance biases, expected from theoretical models (\citealt{Stasinska:1980}). As the metallicity approaches the solar value, strong temperature gradients develop in \hii\/ regions, due to the 
more efficient cooling in the central zone from far-IR \oiii\/ fine-structure lines relative to the outer region. Since the line strength is weigthed towards
warmer regions, due to the strong $T_e$ dependence of the line emissivities,
the net effect of these temperature gradients is that the electron temperature derived from the auroral lines over-estimates the real temperature, and consequently the oxygen abundance is systematically under-estimated (\citealt{Garnett:1992,Stasinska:2005}). In general, in the presence of spatial temperature variations within \hii\/ regions, methods based on collisionally excited lines can lead to under-estimate chemical abundances 
by a factor of two or more (\citealt{Peimbert:1967,Torres-Peimbert:1980}).

Alternative methods of abundance determination, less dependent on the nebular temperature structure than collisionally excited lines, need therefore to be explored in extragalactic \hii\/ regions of high metal content. Metal recombination lines and infrared fine-structure lines offer such alternatives. In both cases the line emissivity is only moderately dependent on $T_e$. 
There are obvious difficulties in exploiting these techniques, however: recombination lines are very weak and hard to detect in extragalactic \hii\/ regions (\citealt{Esteban:2002}), and infrared observations require orbiting telescopes (\citealt{Garnett:2004b}). 

Studies of the O\,{\sc ii} recombination lines in Galactic and extragalactic \hii\/ regions
find oxygen abundances 2 to 3 times larger than those obtained from \oiii\lin4363/(\lin4959\,+\,\lin5007), and conclude that the discrepancy can be explained by the 
presence of temperature fluctuations (\citealt{Peimbert:1993,Peimbert:2003,Garcia-Rojas:2006}). There is still no
general consensus about the source of temperature fluctuations in \hii\/ regions and planetary nebulae (\citealt{Peimbert:2006a}), and alternative explanations for the abundance discrepancy between collisionally excited lines and recombination lines   have been proposed (\citealt{Liu:2000,Tsamis:2005}).

In this paper I report on new deep spectroscopic observations  of four \hii\/ regions in the inner zone of the spiral galaxy M101. The radial oxygen abundance gradient in this galaxy has been studied in detail by previous authors (\citealt{Torres-Peimbert:1989,Kennicutt:2003}, hereafter KBG03). The oxygen abundance of the innermost \hii\/ regions studied thus far, using auroral line detections, is approximately \oh8.7, with an extrapolated central abundance \oh8.76 (KBG03), corresponding to 1.26\,(O/H)$_\odot$. The new observations presented in this paper provide additional high-quality spectroscopic data for some of the most metal-rich \hii\/ regions in this galaxy, with the goals of: $(a)$ obtaining direct abundances closer to the galaxy nucleus from auroral lines; $(b)$ verifying, if feasible, these abundances using metal recombination lines, and $(c)$ measure the effects of temperature fluctuations on the direct abundances.
Observations and data reduction are explained in \S 2. The abundances derived from collisionally excited lines in two \hii\/ regions are presented in \S3, while those derived from recombination lines for helium and carbon for one of them are presented in \S4 and \S5, respectively. A measure of the temperature fluctuations in one of these \hii\/ regions, and their effect on the chemical abundances derived from collisionally excited lines, is given in \S6. In \S7 I discuss the implications of the results obtained for the calibration of strong-line metallicity indicators. The main results of this paper are summarized in \S8.


\section{Observations}

The four targets chosen for this investigation are among the brightest \hii\/ regions in the central 3 arcminutes ($\sim6$~kpc at a distance of 6.85 Mpc, \citealt{Freedman:2001})
of M101: H493, H507 (also Searle~2, from \citealt{Searle:1971}), H972 and H1013 (Searle~3). Table~\ref{journal} contains the identification and the celestial coordinates taken from the catalog of \citet{Hodge:1990}, together with the deprojected galactocentric distance in units of the disk isophotal radius ($R_0$\,=\,14\farcm4), taken from \citet{Kennicutt:1996}.

The spectra of the \hii\/ regions  were obtained on May 12, 2005 with the Keck~I telescope on Mauna Kea using the Low Resolution Imaging Spectrometer (\citealt{oke:1995}). The observing conditions were photometric, and the seeing was stable at 0\farcs7 throughout the night. Using a dichroic splitter, blue and red spectra were obtained simultaneously over most of the optical wavelength range. For the blue spectra (3300-5600~\AA) a 600 lines mm$^{-1}$ grism blazed at 4000~\AA\/ was employed, yielding a spectral resolution of $\sim$5~\AA\/ with a 1.5 arcsec-wide slit. In the red (4960-6670~\AA) a 900 lines mm$^{-1}$ grating blazed at 5500~\AA\/ was used (4~\AA\/ resolution). In order to extend the spectral coverage to the near-infrared \siii\/ lines, spectra were obtained also with a 400 lines mm$^{-1}$ grating blazed at 8900~\AA, covering the 6050-9800~\AA\/ wavelength range at 9~\AA\/ resolution.

The total integration time for the blue and red spectra ranged between 4200 and 4500 seconds, split into two or three exposures. For the near-IR spectra single 300\,s or 900\,s exposures were obtained. I also acquired 300\,s blue and red spectra for the two brightest nebulae, H1013 and H972, in order to avoid saturation of the H$\alpha$ emission line.
Table~\ref{journal} summarizes the exposures secured for this program.

The airmass of M101 varied  during the observations between 1.21 and 1.83. In order to minimize the effects of differential atmospheric refraction the position angle of the slit was adjusted to match the parallactic angle as it varied during the course of the night.
Observations of the standard stars LTT~7987, BD+28d4211 and BD+25d4655 were obtained during twilight for flux calibration.

\begin{deluxetable*}{lcccccc}
\tabletypesize{\scriptsize}
\tablecolumns{7}
\tablewidth{0pt}
\tablecaption{Object sample and exposures acquired\label{journal}}

\tablehead{
\colhead{}	 &
\colhead{R.A.}	&
\colhead{DEC.}	&
\colhead{} &
\multicolumn{3}{c}{Exposures}
        \\
\colhead{\phantom{aaaaaaa}Object\phantom{aaaaaaa}} &
\colhead{(J2000.0)}	&
\colhead{(J2000.0)} &
\colhead{$R/R_0$} &
\colhead{3300-5600~\AA} &
\colhead{4960-6670~\AA} &
\colhead{6050-9800~\AA} }
\startdata
\\[-1mm]
H493\dotfill 	& 14:03:03.4	& +54:21:25		&	 0.10  & 3$\times$1500\,s & 3$\times$1500\,s &   1$\times$900\,s\\[1mm]

H507 (Searle 2)\dotfill	& 14:03:04.2   & +54:19:28		&	 0.14  & 3$\times$1500\,s & 3$\times$1500\,s &   1$\times$900\,s\\[1mm]

H972\dotfill 	& 14:03:27.8   & +54:21:31		&	 0.16  & 2$\times$2100\,s & 2$\times$2100\,s &   1$\times$300\,s\\
		&				&					&	& 1$\times$300\,s & 1$\times$300\,s &      \\[1mm]
		
H1013 (Searle 3)\dotfill	& 14:03:30.7 	& +54:21:14		&	0.19 & 3$\times$1500\,s & 3$\times$1500\,s &   1$\times$300\,s\\
		&				&					&	& 1$\times$300\,s & 1$\times$300\,s &      \\
		
\enddata

\end{deluxetable*}

\subsection{Data reduction and line flux measurement}

The long-slit data reduction was carried out with 
{\sc iraf}\footnote{{\sc iraf} is distributed by the National Optical Astronomy
Observatories, which are operated by the Association of Universities
for Research in Astronomy, Inc., under cooperative agreement with the
National Science Foundation.}, version 2.12, using the {\sc p}y{\sc raf}\footnote{{\sc p}y{\sc raf} is a product of the Space Telescope Science Institute, which is operated by AURA for NASA.} command language. The spectra were corrected for bias, flat-fielded and wavelength calibrated using Hg\,+\,Zn\,+\,Cd lamps (blue range) and Ne\,+\,Ar lamps (red range). For the flux calibration the Mauna Kea extinction curve of \citet{Krisciunas:1987} was employed. Due to the lack of proper calibration data, the flux calibration at wavelengths longer than 9200\,\AA\/ was not carried out.
Spectral regions in common between adjacent wavelength ranges showed a good agreement in the calibrated continuum level and in the flux of lines in common, indicating a 5\% accuracy in the relative flux calibration.

The strength of the emission lines was measured by integration of the flux under the line profile, measured between two continuum points selected interactively. These fluxes were corrected for interstellar reddening adopting the \citet{Howarth:1983} analytical formulation of 
the \citet{Seaton:1979} law. The reddening coefficient C(H$\beta$) was obtained from the Balmer decrement, considering the measured intensities of H$\alpha$ and H$\gamma$ relative to H$\beta$, compared to the case B values taken from \citet{Storey:1995} for an electron temperature $T_e=7500$~K. The equivalent width of the Balmer lines in absorption originating from the underlying stellar component 
was adjusted iteratively in order to reach agreement in the C(H$\beta$) determined separately from the
H$\alpha$/H$\beta$ and H$\gamma$/H$\beta$ ratios. Higher order lines of the Balmer series are progressively more sensitive to the amount of underlying absorption, and were thus not included in this procedure.

The result of the line flux measurements, corrected for reddening, is presented in Table~\ref{fluxes}. The quoted uncertainties account for statistical errors, as well as the uncertainty in the flux calibration, the flat-fielding, and the reddening.
I have compared the fluxes of the strongest metal lines with the values published  by \citet{Kennicutt:1996}. In general there is agreement at the 1-$\sigma$ level. The comparison is slightly worse (2-$\sigma$) for the \oii\lin 3727 line in H1013 and H507. In similar comparisons  (e.g.~\citealt{Bresolin:2005}) the \oii\lin 3727 
is often found to be the one showing  the largest discrepancy. It is not clear if this is due simply to errors in the line flux measurements and/or reddening corrections, or if variations in the regions sampled by the various slits can also account for the observed differences.

\begin{deluxetable*}{cccccc}
\tabletypesize{\scriptsize}
\tablecolumns{6}
\tablewidth{0pt}
\tablecaption{Reddening-corrected line fluxes\label{fluxes}}

\tablehead{
\colhead{Line}	 &
\colhead{}	&
\colhead{H493}	&
\colhead{H507} &
\colhead{H972} &
\colhead{H1013} }
\startdata
\\[-1mm]
3727 &  \oii   &    84  $\pm$  4     	&     84 $\pm$    4	&    139 $\pm$    7	&   221 $\pm$   11\\ 
3750 &  H\,12      &   \nodata		&    1.8 $\pm$  0.2	&    \nodata		&   3.0 $\pm$  0.2\\ 
3771 &  H\,11      &   \nodata		&    4.7 $\pm$  0.3	&    \nodata		&   3.3 $\pm$  0.2\\ 
3798 &  H\,10      &   6.3  $\pm$  0.3	&    5.3 $\pm$  0.3	&    \nodata		&   4.4 $\pm$  0.2\\ 
3822 &  \hei      &   \nodata		&   \nodata		&   \nodata		&  0.72 $\pm$ 0.06\\ 
3835 &  H\,9       &   8.3  $\pm$  0.4	&    7.1 $\pm$  0.4	&    \nodata		&   7.3 $\pm$  0.3\\ 
3868 &  \neiii &   \nodata		&    \nodata		&    \nodata		&   3.1 $\pm$  0.2\\ 
3969 &  H\,8       &   16.0  $\pm$  0.7	&   12.5 $\pm$  0.6	&   18.1 $\pm$  0.7	&  16.4 $\pm$  0.7\\ 
4026 &  \hei      &   \nodata		&   \nodata		&   \nodata		&  1.54 $\pm$ 0.08\\ 
\phantom{~$\dagger$}4072~$\dagger$ &  \sii   &   \nodata		&   \nodata		&   \nodata		&  1.27 $\pm$ 0.07\\ 
4101 &  H$\delta$       &   26.4  $\pm$  1.1	&   26.4 $\pm$  1.1	&   27.5 $\pm$  1.1	&  26.2 $\pm$  1.1\\ 
4144 &  \hei      &   \nodata		&   \nodata		&   \nodata		&  0.13 $\pm$ 0.03\\ 
4267 &  \cii     &   \nodata		&   \nodata		&   \nodata		&  0.25 $\pm$ 0.03\\ 
4340 &  H$\gamma$       &   46.3  $\pm$  1.8	&   46.4 $\pm$  1.9	&   46.4 $\pm$  1.7	&  46.4 $\pm$  1.9\\ 
\phantom{~$\dagger$}4363~$\dagger$ &  \oiii  &   \nodata		&   \nodata		&   \nodata		&  0.24 $\pm$ 0.03\\ 
4388 &  \hei      &   \nodata		&   \nodata		&   \nodata		&  0.40 $\pm$ 0.05\\ 
4471 &  \hei      &   1.2  $\pm$  0.2	&    2.2 $\pm$  0.2	&    1.2 $\pm$  0.1	&   3.9 $\pm$  0.2\\ 
4658 &  \feiii &   \nodata		&   \nodata		&   \nodata		&  0.40 $\pm$ 0.05\\ 
4711 &  \ariv  &   \nodata		&   \nodata		&   \nodata		&  0.35 $\pm$ 0.04\\ 
4861 &  H$\beta$       &   100  $\pm$    4	&    100 $\pm$    4	&    100 $\pm$    4	&   100 $\pm$    4\\ 
4922 &  \hei      &   \nodata		&   1.46 $\pm$ 0.15	&   \nodata		&  1.02 $\pm$ 0.05\\ 
4959 &  \oiii  &     3  $\pm$    0	&      2 $\pm$    0	&      9 $\pm$    0	&    33 $\pm$    1\\ 
4986 &  \feiii &   \nodata		&   0.31 $\pm$ 0.14	&   \nodata		&  0.35 $\pm$ 0.04\\ 
5007 &  \oiii  &     8  $\pm$    0	&      8 $\pm$    0	&     30 $\pm$    1	&   102 $\pm$    4\\ 
5016 &  \hei      &   0.90  $\pm$  0.10	&   2.87 $\pm$ 0.20	&   1.16 $\pm$ 0.15	&  2.44 $\pm$ 0.10\\ 
5041 &  \silii    &   \nodata		&   \nodata		&   \nodata		&  0.17 $\pm$ 0.02\\ 
5048 &  \hei      &   \nodata		&   \nodata		&   \nodata		&  0.13 $\pm$ 0.02\\ 
5056 &  \silii    &   0.55  $\pm$  0.10	&   \nodata		&   \nodata		&  0.17 $\pm$ 0.02\\ 
5199 &  \none    &   1.43  $\pm$  0.13	&   3.17 $\pm$ 0.20	&   0.84 $\pm$ 0.13	&  0.62 $\pm$ 0.03\\ 
5270 &  \feiii &   \nodata		&   \nodata		&   \nodata		&  0.15 $\pm$ 0.02\\ 
5518 &  \cliii &   \nodata		&   \nodata		&   \nodata		&  0.46 $\pm$ 0.03\\ 
5538 &  \cliii &   \nodata		&   \nodata		&   \nodata		&  0.29 $\pm$ 0.02\\ 
\phantom{~$\dagger$}5755~$\dagger$ &  \nii   &   0.24  $\pm$  0.06	&   0.51 $\pm$ 0.12	&   \nodata		&  0.59 $\pm$ 0.03\\ 
5876 &  \hei      &   6.7  $\pm$  0.3	&    6.9 $\pm$  0.3	&    8.5 $\pm$  0.4	&  12.1 $\pm$  0.6\\ 
5978 &  \silii    &   \nodata		&   \nodata		&   \nodata		&  0.13 $\pm$ 0.02\\ 
6046 &  \oi      &   \nodata		&   \nodata		&   \nodata		&  0.09 $\pm$ 0.02\\ 
6300 &  \oi    &   1.17  $\pm$  0.10	&   1.84 $\pm$ 0.11	&   \nodata		&  1.21 $\pm$ 0.07\\ 
\phantom{~$\dagger$}6312~$\dagger$ &  \siii  &   0.27  $\pm$  0.07	&   \nodata		&   \nodata		&  0.77 $\pm$ 0.05\\ 
6360 &  \oi    &   0.37  $\pm$  0.06	&   0.60 $\pm$ 0.06	&   \nodata		&  0.30 $\pm$ 0.03\\ 
6548 &  \nii   &    32  $\pm$    2	&     30 $\pm$    2	&     27 $\pm$    1	&    24 $\pm$    1\\ 
6563 &  H$\alpha$       &   294  $\pm$   15	&    294 $\pm$   16	&    293 $\pm$   15	&   293 $\pm$   16\\ 
6583 &  \nii   &    95  $\pm$    5	&     88 $\pm$    5	&     80 $\pm$    4	&    71 $\pm$    4\\ 
6678 &  \hei      &   2.2  $\pm$  0.2	&    2.3 $\pm$  0.2	&    2.3 $\pm$  0.2	&   3.4 $\pm$  0.2\\ 
6717 &  \sii   &   30.2  $\pm$  1.7	&   30.1 $\pm$  1.7	&   24.1 $\pm$  1.3	&  17.8 $\pm$  1.0\\ 
6731 &  \sii   &   21.2  $\pm$  1.2	&   22.0 $\pm$  1.2	&   16.9 $\pm$  0.9	&  12.0 $\pm$  0.7\\ 
7065 &  \hei      &   \nodata		&   1.49 $\pm$ 0.19	&   1.41 $\pm$ 0.18	&  1.81 $\pm$ 0.11\\ 
7135 &  \ariii &   1.5  $\pm$  0.2	&    1.3 $\pm$  0.1	&    3.8 $\pm$  0.3	&   7.1 $\pm$  0.4\\ 
7236 &  \cii     &   \nodata		&   1.02 $\pm$ 0.18	&   \nodata		&  0.27 $\pm$ 0.03\\ 
7281 &  \hei      &   \nodata		&   \nodata		&   \nodata		&  0.51 $\pm$ 0.05\\ 
\phantom{~$\dagger$}7325~$\dagger$ &  \oii   &   \nodata		&    0.9 $\pm$  0.2	&    \nodata		&   2.4 $\pm$  0.2\\ 
7751 &  \ariii &   \nodata		&   \nodata		&   \nodata		&  1.54 $\pm$ 0.11\\ 
8545 &  P\,15      &   \nodata		&   \nodata		&   \nodata		&  0.69 $\pm$ 0.06\\ 
8598 &  P\,14      &   \nodata		&   \nodata		&   \nodata		&  0.66 $\pm$ 0.07\\ 
8665 &  P\,13      &   \nodata		&   \nodata		&   \nodata		&  0.82 $\pm$ 0.09\\ 
8750 &  P\,12      &   \nodata		&   \nodata		&   \nodata		&  1.07 $\pm$ 0.11\\ 
8863 &  P\,11      &   \nodata		&   1.14 $\pm$ 0.13	&   \nodata		&  1.35 $\pm$ 0.12\\ 
9015 &  P\,10      &   1.51  $\pm$  0.24	&   1.39 $\pm$ 0.15	&   \nodata		&  1.86 $\pm$ 0.15\\ 
9069 &  \siii  &    10  $\pm$    2	&      9 $\pm$    2	&     17 $\pm$    3	&    24 $\pm$    5\\ 
C(H$\beta$) &          &   0.48  $\pm$  0.05	&   0.74 $\pm$ 0.05	&   0.40 $\pm$ 0.05	&  0.35 $\pm$ 0.05\\ 
EW(H$\beta$) &         &    59  $\pm$    1	&     48 $\pm$    1	&     34 $\pm$    1	&   146 $\pm$    1\\ 
F(H$\beta$)\tablenotemark{a}  &         &   29.0  $\pm$    1	&   63.4 $\pm$    1	&   31.8 $\pm$    1	& 157.0 $\pm$    1\\
\enddata
\tablenotetext{a}{Extinction-corrected total flux, in units of 10$^{-15}$ erg\,s$^{-1}$\,cm$^{-2}$.}
\tablecomments{Some lines belonging to a doublet/multiplet are identified by a single wavelength, for example \oii\llin3726,\,3729 (here \oii\lin3727), \sii\llin4069,\,4076 (\sii\lin4072) and \oii\llin7319,7320\,+\,\llin7329,\,7330 (\oii\lin7325). Auroral lines are identified by the $\dagger$ symbol.}
\end{deluxetable*}

As Table~\ref{fluxes} shows, auroral lines have been detected in three of the four targets: H493, H507 and H1013. These lines will be used in Section~3 to derive electron temperatures and ionic abundances from the collisionally excited lines. The \nii\lin5755, \siii\lin6312 and \oii\lin7325 auroral lines have already been measured in H1013 by KBG03. For these lines the comparison with the fluxes presented here is excellent.

\begin{figure*}
\plotone{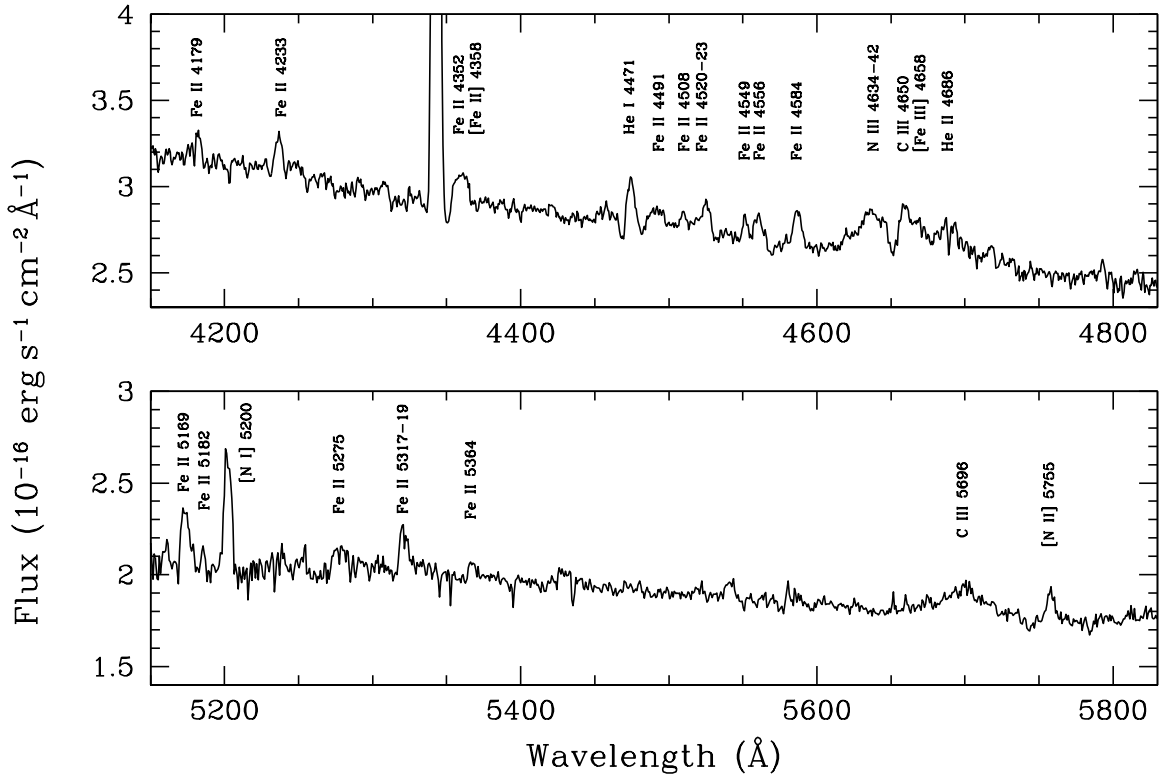}
\caption{The spectrum of H507 contains a large number of \feii\/ lines, some are  shown here in the 4150-4830~\AA\/ (top) and 5150-5830~\AA\/ (bottom) wavelength ranges. The Wolf-Rayet features at $\sim$\,4650~\AA\/ and 5696~\AA\/ are also labeled.\label{h507}}
\end{figure*}

The spectrum of H507 shows a large number of \feii\/ lines (Fig.~\ref{h507}), resembling the emission-line rich spectra of known luminous Galactic iron stars, such as HD~316285 and $\eta$ Carinae (\citealt{Walborn:2000}). These are massive stars believed to be in an advanced stage of evolution.
The presence of evolved massive stars in H507 is also evident from the Wolf-Rayet emission features at $\sim$\,4650\,\AA\/ (both the {\sc n\,iii}\,\llin 4634-42 and the \heii\lin4686 are detected, indicative of WN stars) and 5696\,\AA\/ from {\sc c\,iii} (late WC stars). In fact, W-R emission lines have been detected in all four \hii\/ regions studied here. The high detection fraction of W-R features, often including the signatures of WC stars, in high-metallicity \hii\/ regions has been noted in earlier works (\citealt{Bresolin:2004,Bresolin:2005,Hadfield:2005}), and is expected from the increase in the stellar mass loss with 
metallicity (\citealt{Meynet:2005}). Given the scope of this paper, I will not further discuss the W-R content of the \hii\/ regions under consideration, and I will
simply note that
the stellar envelope of stars with \feii\/ lines in emission can also be responsible for emission in auroral lines such as \nii\lin5755 and \oii\lin7325 (\citealt{Hillier:1998}, \citealt{Borges-Fernandes:2001}). Therefore, these lines will not be used to measure nebular electron temperatures in H507.

Electron densities were estimated from the \sii\lin6716/\lin6731 line ratio,
which in all \hii\/ regions is consistent with the low-density limit. This is also the case for the \cliii\lin5518/\lin5538 ratio measured for H1013, compared to the theoretical low-density value (\citealt{Keenan:2000}). A common density $\rm N_e = 100~cm^{-2}$ will therefore be adopted in the remainder of the paper for all \hii\/ regions.


\section{Metal abundances from collisionally excited lines}

The following analysis focuses on the two \hii\/ regions for which reliable electron temperatures could be measured: H493 and H1013. 
For the derivation of ionic abundances from collisionally excited lines I relied on the programs by \cite{Shaw:1995} available in the {\sc stsdas} {\em nebular} 
package running under {\sc iraf}. The atomic parameters and the collision strengths adopted are those implemented in the 1997 version of the software, except for the {\sc s\,iii} collision strengths, which were taken from 
\citet{Tayal:1999}.

In the case of H1013 five values of the electron temperature were derived from the ratios of the available auroral lines to the corresponding nebular lines (see Table~\ref{te}). The values from the different ions are all similar to each other, and are in the 7680--8360~K range.
The ionic abundances were derived adopting $T$\oiii\/ for O$^{++}$ and Ne$^{++}$, $T$\siii\/ for S$^{++}$ and Ar$^{++}$, $T$\sii\/ for S$^+$, and a weighted average of $T$\oii\/ and $T$\nii\/, corresponding to $8070\pm120$~K, for O$^+$ and N$^+$.

For H493 only $T$\siii\/ and $T$\nii\/ could be directly measured. In this case, therefore, I determined the temperature relative to the O$^{++}$ emission zone (assumed to be equal to the temperature measured from the \oiii\/ lines) following previous papers in this series (\citealt{Bresolin:2004, Bresolin:2005}), i.e.~by adopting the relationship between $T$\oiii\/ and $T$\siii\/ predicted by photoionization models (\citealt{Garnett:1992}, \citealt{Stasinska:1982}):

\begin{equation}
{\rm {\it T}[S\,{\scriptstyle III}] = 0.83~{\it T}[O\,{\scriptstyle III}]~+~1700~K}
\end{equation}

\noindent
where the electron temperatures are in K. However, because oxygen is present in H493 almost exclusively as O$^+$, this procedure has a very little impact on the derived total oxygen abundance. The other assumption made in the abundance analysis of H493 is that $T$\nii\/ can be used for the determination of ionic abundances of all the low-ionization species.

\begin{deluxetable}{lccc}
\tabletypesize{\scriptsize}
\tablecolumns{4}
\tablewidth{0pt}
\tablecaption{Measured electron temperatures (K)\label{te}}

\tablehead{
\colhead{\phantom{aaaaaaaaaaaaaa}}	 &
\colhead{$\rm\frac{auroral}{nebular}$} &
\colhead{H493}	&
\colhead{H1013}}
\startdata
\\[-1mm]
T\oiii\dotfill	& $\frac{4363}{4959+5007}$  &  \nodata &  7700 $\pm$ 250 \\[2mm]
T\oii\dotfill	& $\frac{7325}{3727}$       &  \nodata &  7690 $\pm$ 200 \\[2mm]
T\nii\dotfill	& $\frac{5755}{6548+6583}$  &  5960 $\pm$ 340  &  8360 $\pm$ 150 \\[2mm]
T\sii\dotfill   & $\frac{4072}{6717+6731}$  &  \nodata &  8160 $\pm$ 340 \\[2mm]
T\siii\dotfill	& $\frac{6312}{9069+9532}$  &  7330 $\pm$ 590  &  7680 $\pm$ 160 \\
\enddata

\end{deluxetable}

\begin{deluxetable}{lcc}
\tabletypesize{\scriptsize}
\tablecolumns{3}
\tablewidth{0pt}
\tablecaption{Ionic and total abundances from collisionally excited lines\label{ionic}}

\tablehead{
\colhead{\phantom{aaaaaaaaaaaaaa}}	 &
\colhead{H493}	&
\colhead{H1013}}
\startdata
\\[-1mm]
$\rm O^{+}/H^+$\dotfill	  &	$5.4 \pm 2.3 \times10^{-4}$   & $2.3\pm0.2 \times10^{-4}$  \\
$\rm O^{++}/H^+$\dotfill  &	$1.4 \pm 0.8 \times10^{-5}$   & $9.9\pm1.4 \times10^{-5}$  \\
$\rm N^{+}/H^+$\dotfill	  &	$1.0 \pm 0.2 \times10^{-4}$   & $2.6\pm0.1 \times10^{-5}$  \\
$\rm S^{+}/H^+$\dotfill	  &	$6.1 \pm 1.4 \times10^{-6}$   & $1.2\pm0.1 \times10^{-6}$  \\
$\rm S^{++}/H^+$\dotfill  &	$3.1 \pm 0.7 \times10^{-6}$   & $6.7\pm0.3 \times10^{-6}$  \\
$\rm Ar^{++}/H^+$\dotfill & $3.3 \pm 0.8 \times10^{-7}$   & $1.3\pm0.1 \times10^{-6}$  \\
$\rm Ne^{++}/H^+$\dotfill & \nodata                       & $1.1\pm0.2 \times10^{-5}$  \\[2mm]
12\,+\,log(O/H)\dotfill       & $8.74 \pm 0.17$               & $8.52 \pm 0.05$ \\
log(N/O)\dotfill          & $-0.72 \pm 0.22$              & $-0.95 \pm 0.05$ \\
log(S/O)\dotfill          & $-1.78 \pm 0.20$              & $-1.61 \pm 0.05$ \\

\enddata

\end{deluxetable}

From the electron temperatures in Table~\ref{te} the ionic abundances presented in Table~\ref{ionic} were derived. To calculate the total abundances of oxygen, nitrogen and sulfur, the following equations were used:

\begin{equation}
\rm O/H = (O^+ + O^{++}) / H^+
\end{equation}

\begin{equation}
\rm N/O = N^+/O^+
\end{equation}

\noindent
as suggested by \citet{Peimbert:1969}, and

\begin{equation}
\frac{{\rm S}^+ + {\rm S}^{++}}{{\rm S}} = \left[ 1 - \left(
  1-\frac{{\rm O}^+}{{\rm O}} \right)^\alpha \right] ^{1/\alpha},
\end{equation}

\noindent
following \citet{Stasinska:1978} with $\alpha=2.5$. This last equation yields very small corrections ($\sim 1\%$ or less) to the sulfur abundance in both H493 and H1013, due to the low amount of 
$\rm S^{3+}$ expected in low-excitation nebulae (see also KBG03, \citealt{Bresolin:2004}). The O/H, N/O and S/O abundance ratios thus derived are included in Table~\ref{ionic}.

KBG03 determined the oxygen abundance gradient in M101 from
20 \hii\/ regions in which the electron temperature of the gas could be determined from ratios of auroral lines to nebular lines. The galactocentric distances of these \hii\/ regions vary between $R/R_0=0.19$ (H1013) and $R/R_0=1.25$ (SDH323).
The new measurements presented here add one further point closer to the center of the galaxy, H493 at $R/R_0=0.10$, and allow a determination of the $\rm O^{++}$ abundance of H1013 from a direct measurement of \oiii\lin4363. The resulting O/H abundance is $\sim$0.2 dex lower than measured by KBG03: $\rm 12+log(O/H)=8.52\pm0.06$ (this work) vs.~$\rm 12+log(O/H)=8.71\pm0.05$ (KBG03). 
The differences in the adopted O$^+$ temperature (8070~K vs.~7600~K)
and in the O$^{++}$ temperature (7700~K vs.~6600~K) are responsible, in approximately equal proportions, for the
discrepancy in abundance. The $\rm O^{++}$ zone temperature was
derived by KBG03 from $T$\siii, measured from the \siii\lin6312/(\lin9069\,+\,\lin9532) line ratio, and the application of Equation (1). The $T$\siii\/ value determined in the current work is $\sim$500~K higher than the KBG03 value, as a result of a $\sim$35\% smaller flux in the \siii\llin9069,\,9532 lines.
I point out that Equation (1), together with the value of $T$\siii\/ in Table~\ref{te}, would provide $T$\oiii\,=\,$7200\pm200$~K, i.e.~500~K smaller than measured directly
from \oiii\/\,\lin4363/(\lin4959\,+\,\lin5007). The relation proposed by \citet{Perez-Montero:2005}, ${\rm {\it T}[S\,{\scriptstyle III}] = 1.05~{\it T}[O\,{\scriptstyle III}]~-~800~K}$, would yield instead $T$\oiii\,=\,$8100\pm150$~K.
This result is a warning that relations derived from photoionization models, such as Equation (1), can work well in a statistical sense, but can fail for individual \hii\/ regions when uncertainties in \te\/
smaller than a few hundred degrees are required.

A weighted linear least square fit to the KBG03 data points, supplemented by the abundances determined here for H409 and H1013, provides the following solution for the radial oxygen abundance gradient in M101:

\begin{equation}
\rm
12 + log(O/H) = 8.75\, (\pm 0.05)\, - \,0.90\, (\pm 0.07)\, R/R_0.
\end{equation}

\begin{figure}
\plotone{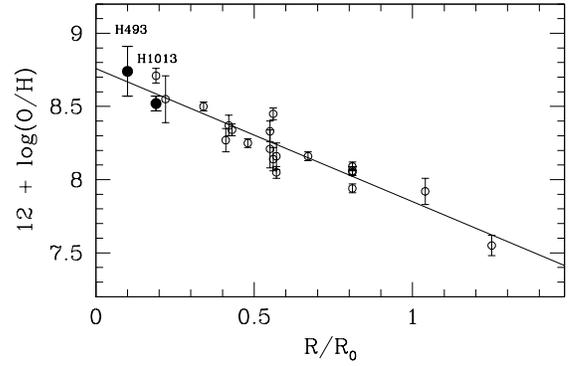}
\caption{The oxygen abundance gradient in M101 in terms of the fractional 
isophotal radius, $R/R_0$, determined from the KBG03 measurements (open circles) and the 
inner regions H493 and H1013 studied here (full dots). All the data points 
represent \hii\/ regions where the electron temperature has been measured
directly from auroral-to-nebular line ratios. In all cases, except for the innermost region, H493, and H336 (=\,Searle~5, point at $R/R_0=0.22$) the \oiii\/ temperature has been derived from the \lin 4363 line. In this plot
the straight line represents the
weighted linear least square fit, given by Eq.~(5).
\label{radial}}
\end{figure}

\noindent
This relation is virtually the same as the one determined by KBG03. The innermost region studied in this paper, H493, 
allows us to extend the empirical abundance gradient in M101 to 1.5 arcmin from the galactic center. The oxygen abundance of H493, $\rm 12+log(O/H)=8.74\pm0.17$,
agrees well with the galactic abundance gradient, as shown in Fig.~\ref{radial}.

\section{Helium abundances}

There are several \hei\/ recombination lines which could, in principle, be used to obtain the helium abundance, especially in the case of H1013 (see Table~2). However, many of these lines are faint, with small equivalent widths, and the abundances derived from them would be quite sensitive to even small corrections for underlying absorption. Some of the stronger lines are not well suited for the measurement of the helium abundance either, because they are partially blended with strong metal lines, as in the case of \hei\lin5016 with \oiii\lin5007, or they can be strongly affected by optical depth effects (\hei\lin7065). Therefore, only the two lines \hei\lin5876 and \lin6678 have been used.

The $\rm He^+/H^+$ ionic ratios have been obtained from  \hei\lin5876/H$\beta$ and \hei\lin6678/H$\beta$, adopting the emissivities of \citet{Benjamin:1999} for \hei\/ and of \citet{Storey:1995} for \hi, both calculated at the $T$\oiii\/ temperature determined in the previous section. The average ionic abundances thus derived from \lin5876 and \lin6678 are reported in Table~\ref{recomb}. In the case of
the low excitation \hii\/ region H493 I find $\rm He^+/H^+ = 0.048$, almost half the value found in H1013, $\rm He^+/H^+ = 0.083$. These helium ionic fractions should be compared to oxygen ionic fractions  $\rm O^+/(O^+ + O^{++})$ of 0.97 (H493) and 0.70 (H1013; a value of 0.55 is however derived when accounting for the presence of temperature fluctuations, as shown in \S~6). 

A large fraction of the helium present in low-excitation nebulae is in the neutral form. To estimate its contribution to the total helium abundance, a grid of photoionization models has been obtained with {\sc cloudy} 94.00 (\citealt{Ferland:1998}). The helium ionization correction factor, ICF(He), has been estimated from the predicted relation between $\rm He^0/(He^0 + He^+)$ and $\rm O^+/(O^+ + O^{++})$ ($\rm He^{++}$ is negligible at these low excitation levels). I found $\rm ICF(He)=2.13\pm0.23$ for H493 and 
$\rm ICF(He)=1.28\pm0.09$ for H1013 ($1.20\pm0.09$ accounting for temperature fluctuations).
The resulting helium abundances for H493 ($\rm He/H=0.10\pm0.01$) and H1013 [$\rm He/H=0.11\pm0.01$, or $\rm He/H=0.10\pm0.01$ if using $\rm O^+/(O^+ + O^{++})$\,=\,0.55] are in agreement, within the uncertainties. 

\begin{deluxetable}{lcc}
\tabletypesize{\scriptsize}
\tablecolumns{3}
\tablewidth{0pt}
\tablecaption{Ionic and total abundances from recombination lines\label{recomb}}

\tablehead{
\colhead{\phantom{aaaaaaaaaaaaaa}}	 &
\colhead{H493}	&
\colhead{H1013}}
\startdata
\\[-1mm]
$\rm He^+/H^+$\dotfill        & $0.048 \pm 0.003$             & $0.083 \pm 0.001$\\
$\rm He/H$\dotfill            & $0.10 \pm 0.01$               & $0.10 \pm 0.01$\\[2mm]
$\rm C^{++}/H^+$\dotfill      & \nodata                       & $2.3 \pm 0.3 \times 10^{-4}$\\
$\rm 12+log(C^{++}/H^+)$      & \nodata                       & $8.36 \pm 0.06$\\
$\rm 12+log(C/H)$             & \nodata                       & $8.66 \pm 0.07$\\
\enddata

\end{deluxetable}


\section{The carbon abundance of H1013}

\begin{figure*}
\plotone{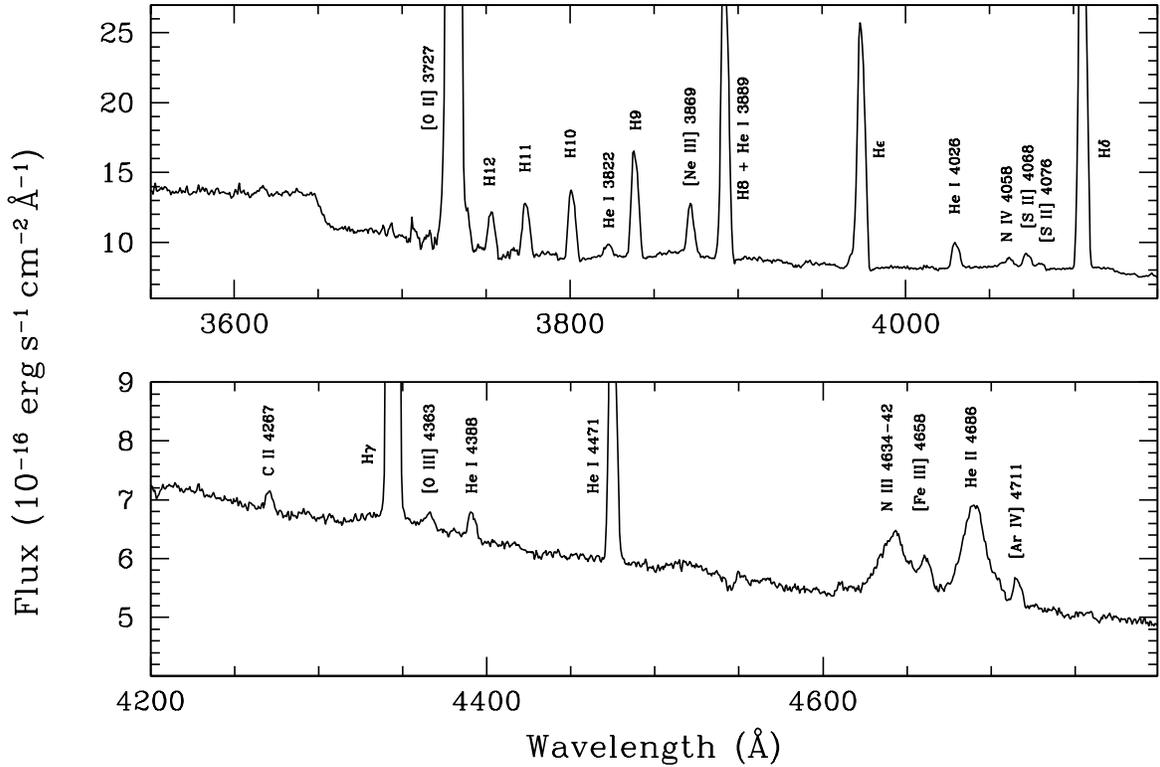}
\caption{Portions of the spectrum of H1013 showing the Balmer discontinuity region and the high-order Balmer lines (top), and the \cii\lin4267 recombination line, the \oiii\lin4363 auroral line and the Wolf-Rayet broad emission features around 4650\AA\/ (bottom).
\label{h1013}}
\end{figure*}

The \cii\lin4267 recombination line has been detected in H1013, the brightest of the \hii\/ regions studied in this work (Fig.~\ref{h1013}). \citet{Esteban:2002} measured this line, as well as various O\,{\sc ii} lines, in two \hii\/ regions of lower metallicity in M101, NGC~5471 and NGC~5461. Metal recombination lines allow the determination of ionic abundances that are virtually independent of the electron temperature, thanks to the fact that they have a similar temperature dependence as the hydrogen recombination lines, and therefore allow very important checks on the abundances obtained from collisionally excited lines. Unfortunately, no O\,{\sc ii} lines could be measured in the LRIS spectra. This is likely a result of the low spectral resolution, combined with the fact that the strongest oxygen recombination lines around 4650~\AA\/ fall in the spectral region occupied by the strong W-R blue bump. Besides, at the low excitation levels typical of the \hii\/ regions considered here, most of the oxygen is in the $\rm O^+$ form rather than $\rm O^{++}$, as shown in the previous section. As a consequence, the O\,{\sc ii} recombination lines will be very faint. This problem affects most metal-rich \hii\/ regions, which are generally of low excitation (see Sect.~7).

Because of the lack of collisionally excited lines from carbon in the optical spectra of \hii\/ regions, and the current unavailability of UV spectrographs working in space for the study of lines like C\,{\sc iii}]\lin1909, the derivation of the $\rm C^{++}/H^+$ ratio and of the total C/H abundance in extragalactic nebulae from the detection of recombination lines is very important to study the chemical evolution of spiral galaxies. In the case of the center of M101, it is interesting to  test whether the  \cii\lin4267 recombination line provides a carbon abundance that is consistent with the measured oxygen abundance. For this purpose, I have adopted the \cii\/ effective recombination coefficient $\rm \alpha_{C\,II}$ from \citet{Davey:2000} and used the following expression to derive the abundance of $\rm C^{++}$:

\begin{equation}
\rm \frac{C^{++}}{H^+} = \frac{I(4267)}{I(H\beta)}\,\frac{4267}{4861}\,\frac{\alpha_{H\beta}}{\alpha_{C\,II}}
\end{equation}

\noindent
where the $\rm H\beta$ effective recombination coefficient $\alpha_{H\beta}$ has been obtained from the \citet{Storey:1995} emissivities. Both $\alpha$'s have been calculated at the $T$\oiii\/ temperature. The resulting ionic abundance fraction is 
$\rm C^{++}/H^+ = 2.3\times10^{-4}$, equivalent to $\rm 12+log(C^{++}/H^+)=8.36\pm0.06$ (Table~\ref{recomb}).

\begin{figure}
\plotone{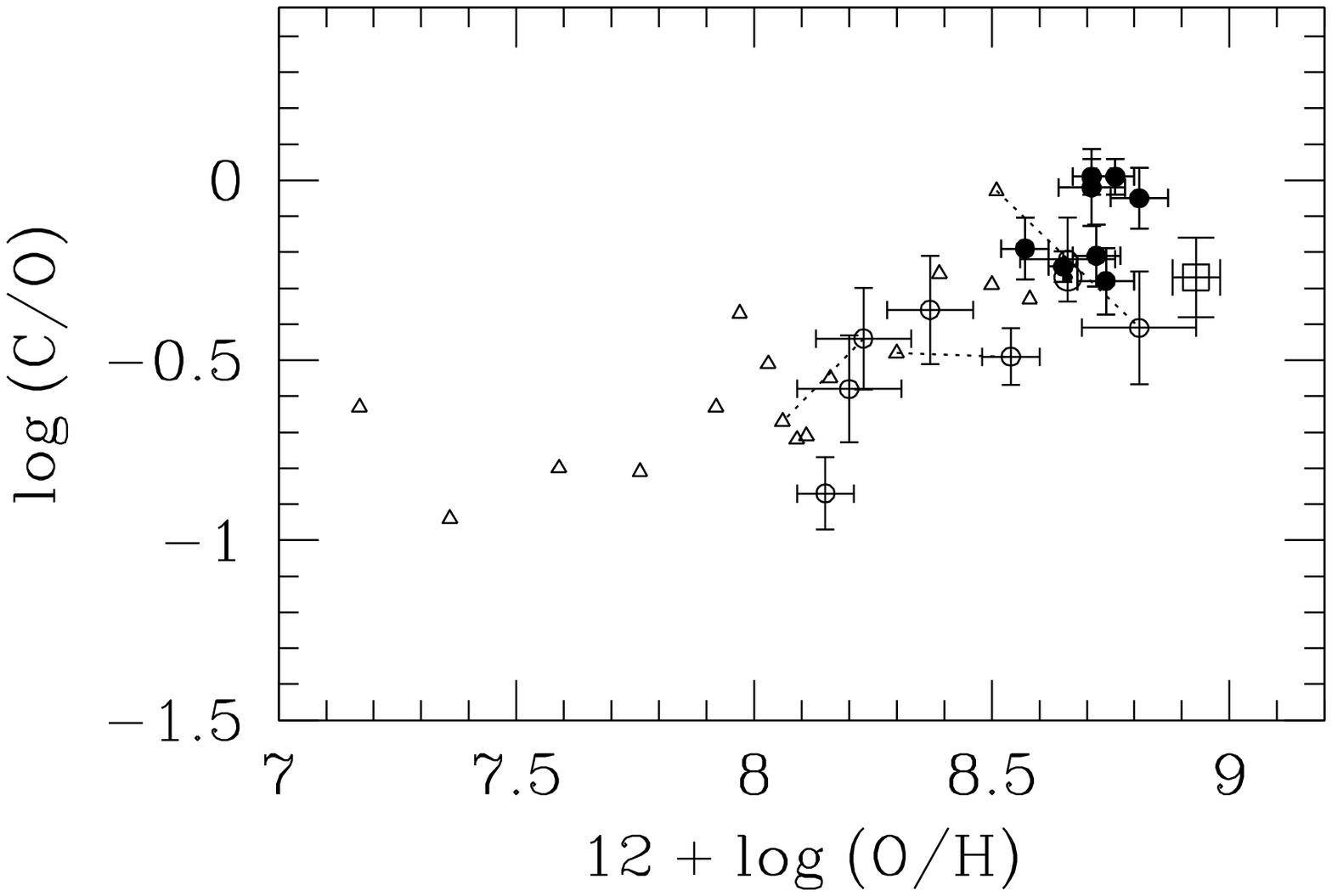}
\caption{C/O ratio as a function of the oxygen abundance O/H from different samples of \hii\/ regions. Small triangles represent nebulae in spirals and irregulars for which the abundances have been obtained from collisionally excited lines  (\citealt{Garnett:1995,Garnett:1997,Garnett:1999,Kobulnicky:1998}). Recombination line abundances exist instead for the 
objects represented by open (extragalactic) and solid (Galactic) circles (\citealt{Esteban:2002,Esteban:2005,Peimbert:2003,Peimbert:2005}). The dotted lines connect \hii\/ regions in common between the two sets. The H1013 \hii\/ region is 
shown as a large open square.
\label{co}}
\end{figure}

In order to account for the $\rm C^+$ contribution to the total C abundance ($\rm C^{3+}$ is negligible in the low-excitation regime), a procedure similar to the one adopted for the helium ICF has been followed, i.e.~the $\rm C^+/C$ fraction as a function of $\rm O^+/O$ predicted by photoionization models was considered. The result
$\rm C^+/C=0.61 \pm 0.02$ has been obtained. The models by \citet{Garnett:1999} would yield the same result. 
Adopting instead $\rm O^+/O\,=\,0.55$, as determined in \S~6, I obtain instead
$\rm C^+/C=0.50 \pm 0.02$. The latter value is adopted, to infer a 
total carbon abundance for H1013 of $\rm 12+log(C/H)=8.66\pm0.06$. This value corresponds to 1.9 times the solar value (\citealt{Asplund:2005}). Fig.~\ref{co} compares the O/H and C/O abundance determined for H1013 with 
data from the literature. Abundances obtained from collisionally excited lines (\hii\/ regions in spiral and irregular galaxies from \citealt{Garnett:1995,Garnett:1997,Garnett:1999,Kobulnicky:1998}) and from recombination lines (\hii\/ regions in the Milky Way and other spirals, \citealt{Esteban:2002,Esteban:2005,Peimbert:2003,Peimbert:2005}) are shown. The H1013 datapoint should be compared with the latter set only, but the collisionally excited line results are also included for completeness (the dotted lines connect \hii\/ regions in common between the two sets: NGC~5461 and NGC~5471 in M101, and 30 Doradus in the LMC).

\section{Temperature fluctuations in H1013}

Thanks to the excellent blue sensitivity of LRIS, the Balmer discontinuity at 3646\,\AA\/  is very well observed in the nebular spectra presented in this work (the case of H1013 can be seen in the top panel of Fig.~\ref{h1013}). The nebular Balmer jump offers the opportunity to derive an electron temperature that is independent of the collisionally excited lines of metals, and representative of the hydrogen continuum spectrum, $T(Bac)$. As shown originally by \citet{Peimbert:1967} and \citet{Peimbert:1969}, the comparison between two independently derived temperatures, e.g.~$T(Bac)$ and $T$\oiii, provides an indication about the presence of temperature fluctuations in \hii\/ regions and planetary nebulae. In the following 
I analyze the \hii\/ region H1013, where the contribution to the
nebular continuum by underlying stars (direct and scattered) appears to be negligible, as judged by the small equivalent width (0.1~\AA) of the Balmer lines in absorption inferred during the reddening correction. I follow the technique developed by \citet{Peimbert:1969} and summarized in several recent papers, for example by \citet{Peimbert:2004}.

As a first step, the electron temperature from the Balmer jump has been derived. 
For this purpose, one needs to reproduce the observed Balmer discontinuity, $I(3646^-)-I(3646^+)$, accounting for the  continuum processes at work: free-free and free-bound transitions of \hi, \hei\/ and \heii, and two-photon decay of the $2^2S_{1/2}$ level of hydrogen. The corresponding temperature-dependent coefficients have been taken from \citet{Brown:1970}, neglecting the \heii\/ contribution, because of the small excitation of H1013, and adopting the $\rm He^+/H^+$ ratio in Table~\ref{recomb} for the calculation of the \hei\/ contribution. 
The spectral resolution of the H1013 spectrum is not sufficient to allow a measurement of the real continuum in proximity of the Balmer discontinuity, because the blending of the high-order Balmer lines creates a raised pseudo-continuum. I have therefore extrapolated a fit to the nebular continuum in the $\sim\,3800-3950$~\AA\/ wavelength range to the wavelength of the Balmer discontinuity. Using then the ratio of the discontinuity to the H$\beta$ and H$\gamma$ emission lines I have derived $T(Bac)=5000\pm800$~K, where the error is dominated by the extrapolation of the continuum.

In order to derive the average temperature $T_0$ and the mean square temperature fluctuation $t^2$ the following equations were used:

\begin{equation}
T{\rm [{\scriptstyle OIII}]} = T_0 \left[ 1 + \left(\frac{91300}{T_0}-3 \right)\frac{t^2}{2} \right]
\end{equation}

\begin{equation}
T(Bac) = T_0 (1 - 1.67\,t^2)
\end{equation}

\noindent
yielding $T_0 = 5500\pm700$~K, $t^2 = 0.06\pm0.02$. Comparable values of  $t^2$ have been found for other extragalactic \hii\/ regions. For example, 
$t^2=0.041$ in NGC~5461  and $t^2=0.074$ in NGC~5471 (both \hii\/ regions located in M101 and studied by \citealt{Esteban:2002}), $t^2=0.064-0.098$ in NGC~2363 (\citealt{Gonzalez-Delgado:1994}), $t^2=0.054-0.076$ in regions X and V of NGC~6822 (\citealt{Peimbert:2005}), $t^2=0.033$ in 30 Doradus (\citealt{Peimbert:2003}).

The effect of temperature fluctuations is to increase the chemical abundance measured from collisionally excited lines determined assuming $t^2=0$.
The ionic abundances in Table~\ref{ionic} were therefore corrected for the case $t^2\neq0$, following \citet{Peimbert:1969}. The results are summarized in Table~\ref{corrected}. As the table shows, the correction to the oxygen abundance amounts to about 0.4 dex. This upward corrections to O/H is somewhat larger than typically found comparing results from collisionally excited lines and recombination lines. The average correction in a number of \hii\/ regions studied in the Milky Way is 0.2 dex
(see \citealt{Garcia-Rojas:2006}, \citealt{Peimbert:2005a}, and references therein), ranging between 0.11 dex and 0.25 dex. Slightly larger values are found in the extragalactic \hii\/ regions NGC~5461 (0.25 dex, \citealt{Esteban:2002}) and Hubble~V in NGC~6822 (0.29 dex, \citealt{Peimbert:2005}).

In conclusion, accounting for the presence of temperature fluctuations, the oxygen abundance of H1013 is $\rm 12+log(O/H)=8.93\pm0.06$. Using the carbon abundance from \S~5, the carbon-to-oxygen ratio is $\rm log(C/O)=-0.27 \pm 0.11$. The measurement of the O\,{\sc ii} recombination lines in H1013 would improve the accuracy of these determinations, and will require spectroscopic observations at higher spectral resolution than the one used for the current work.

\begin{deluxetable}{lcc}
\tabletypesize{\scriptsize}
\tablecolumns{3}
\tablewidth{0pt}
\tablecaption{H1013: ionic and total abundances corrected for $t\neq0$\label{corrected}}

\tablehead{
\colhead{\phantom{aaaaaaaaaaaaaaaaaa}}	 &
\colhead{$t^2=0.00$}	&
\colhead{$t^2=0.06$}}
\startdata
\\[-1mm]
$\rm O^{+}/H^+$\dotfill	  &  $2.3\pm0.2 \times10^{-4}$ & $4.7\pm0.4\times10^{-4}$ \\
$\rm O^{++}/H^+$\dotfill   &  $9.9\pm1.4 \times10^{-5}$ & $3.9\pm0.6\times10^{-4}$ \\
$\rm N^{+}/H^+$\dotfill	  &	 $2.6\pm0.1 \times10^{-5}$ & $5.1\pm0.2\times10^{-5}$ \\
$\rm S^{+}/H^+$\dotfill	  &  $1.2\pm0.1 \times10^{-6}$ & $2.4\pm0.2\times10^{-6}$\\
$\rm S^{++}/H^+$\dotfill  &	 $6.7\pm0.3 \times10^{-6}$ & $2.9\pm0.1\times10^{-5}$\\
$\rm Ar^{++}/H^+$\dotfill &   $1.3\pm0.1 \times10^{-6}$ & $4.1\pm0.3\times10^{-6}$\\
$\rm Ne^{++}/H^+$\dotfill &   $1.1\pm0.2 \times10^{-5}$ & $4.7\pm0.8\times10^{-5}$\\[2mm]
12\,+\,log(O/H)\dotfill   &   $8.52 \pm 0.05$           & $8.93 \pm 0.05$\\
log(N/O)\dotfill          & $-0.95 \pm 0.05$              & $-0.96 \pm 0.05$ \\
log(S/O)\dotfill          & $-1.61 \pm 0.05$              & $-1.47 \pm 0.07$ \\
\enddata

\end{deluxetable}


\section{Discussion}
In this section I consider the results obtained for the inner \hii\/ regions of M101 within the context of the calibration of nebular metallicity indicators based on strong lines, in particular $R_{23}$. 
In general most of the following considerations could be applied to other strong-line diagnostics, some of which do not suffer
some of the problems that affect $R_{23}$.
The major one is the well-known double valuedness of the parameter, that is the fact that a given $R_{23}$ value corresponds to two values of the oxygen abundance. An emission line diagnostic that is
monotonically increasing with abundance, such as \nii/\oii, can be used to remove the degeneracy and place an \hii\/ region in the correct branch.
For the metal-rich \hii\/ regions that are of interest here, only the upper branch calibration is relevant.
Secondly, $R_{23}$ is sensitive to the ionization parameter (\citealt{Perez-Montero:2005}). This
can be accounted for, in principle, using \oiii/\oii\/ (\citealt{McGaugh:1991}) or \oiii/(\oiii\,+\,\oii) (the $P$ index, \citealt{Pilyugin:2001}). Despite these drawbacks,
$R_{23}$ is preferentially considered in what follows because of its wide application in the literature, and because much of the recent empirical work deals with this indicator.

\subsection{Abundance biases}

This work has extended the auroral line method of abundance determination to a fractional galactocentric distance in M101 of $R/R_0\,=\,0.10$ ($\sim 3$~kpc) and to an oxygen abundance \oh8.74. The innermost \hii\/ region studied by KBG03 was H1013 ($R/R_0\,=\,0.19$). The oxygen abundance they determined from the \siii\/ and \nii\/ auroral lines has been revised to a $\sim$0.2 dex lower value with the detection of \oiii\lin4363 in this paper. Both H1013 and the \hii\/ region closest to the center analyzed here, H493, have been found to fit the oxygen abundance gradient in M101 quite well. A systematic offset
from this gradient would be expected if metal-rich \hii\/ regions are affected by strong abundance biases.
The result obtained for H493 and H1013 then suggests that the biases due to large-scale temperature fluctuations in \hii\/ regions of approximately solar metallicity, \oh8.7, are not as large as the maximum value predicted by photoionization models (a $\sim$0.2 dex systematic effect in the case of abundances based on \oiii\lin4363, 
\citealt{Stasinska:2005}), or that the expected abundance biases become relevant at a somewhat larger metallicity, perhaps around \oh9.0. In either case, the abundances obtained from auroral lines in this and in previous papers (e.g.~\citealt{Bresolin:2004}) up to approximately the solar metallicity, i.e.~what one finds in the very central regions of metal-rich spiral galaxies, appear to be robust (a similar argument is also presented by \citealt{Pilyugin:2006}).

\subsection{Metallicity dependent excitation}

The direct abundances measured in recent years from auroral-to-nebular line ratios  (\citealt{Castellanos:2002}; KBG03, \citealt{Bresolin:2004,Bresolin:2005}) have provided the necessary input for the empirical calibration of strong-line abundance indicators
for metal-rich \hii\/ regions. The $P$-method of \citet{Pilyugin:2000,Pilyugin:2001} has been recently recalibrated at high metallicity by \citet{Pilyugin:2005} on the basis of the recent \te-based abundances. As mentioned earlier, this empirical calibration leads to \hii\/ region abundances that are, in the case of objects located in the metal-rich, upper branch of $R_{23}$, factors of 2-3 lower than those derived from calibrations based on photoionization models. In order to calibrate indicators that can provide abundances of star-forming galaxies across the universe with accuracies of, say, 0.1 dex, it is crucial to understand the origin of the discrepancies between different calibrations, and try to test them against additional empirical methods.
The main difficulty is to provide accurate metallicities for \hii\/ regions in the $\log R_{23} < 0.5$ regime, corresponding approximately to (O/H)\,$>$\,(O/H)$_\odot$, due to the weakness of the direct abundance diagnostics.

\begin{figure}
\plotone{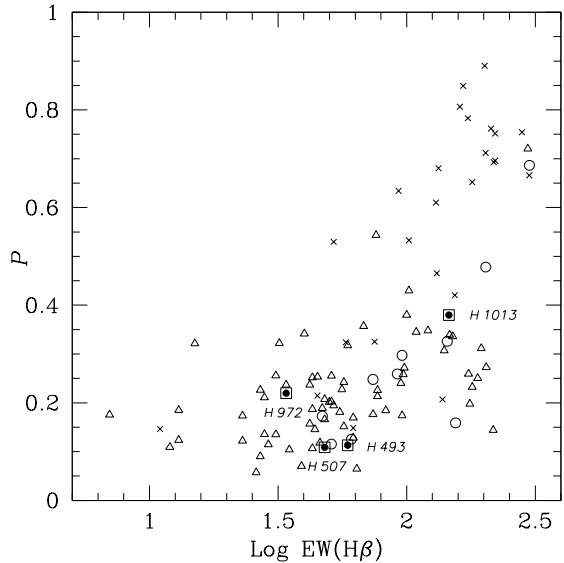}
\caption{Trend of decreasing equivalent width of H$\beta$ with decreasing excitation parameter $P$. The dots represent \hii\/ regions samples in M101 (KBG03, crosses), M51 (\citealt{Bresolin:2004}, open circles), and in five additional spiral galaxies (\citealt{Bresolin:2005}, open triangles). The four M101 \hii\/ regions studied in this work are labeled (square-dot symbols). 
\label{p}}
\end{figure}

In Pilyugin's method, $R_{23}$ depends not only on the oxygen abundance, but also on a measure of the 
nebular ionization, parameterized by the excitation parameter $P$\,=\,\oiii\llin4959,\,5007 / (\oiii\llin4959,\,5007\,+\,\oii\lin3727). This follows the original suggestion by \citet{McGaugh:1991} to introduce a two-dimensional characterization of nebular spectra. It is worth pointing out here the empirical finding that the extragalactic \hii\/ regions that extend along the upper branch of the $R_{23}$ vs.~(O/H) diagram below $\log R_{23}\,=0.5$ tend to have a low excitation parameter. In fact,  an increase in the metallicity (decreasing $R_{23}$) corresponds, in a statistical sense, to a decrease in $P$, i.e.~high metallicity \hii\/ regions with a measured value of \te\/ tend to be also low excitation nebulae. This can be seen by plotting $R_{23}$ against O/H, together with curves of constant $P$ (see, for example, Fig.~12 of \citealt{Pilyugin:2005}), and can be
viewed in light of the 
 correlation between galactocentric distance and equivalent width of the H$\beta$ nebular emission line, EW(H$\beta$), noted in the early abundance studies of spiral galaxies (\citealt{Searle:1971}, \citealt{Shields:1976}). This is shown in Fig.~\ref{p}, where log\,EW(H$\beta$) is plotted against $P$ for the \hii\/ region samples of KBG03, \citet{Bresolin:2004,Bresolin:2005} and for the M101 \hii\/ regions studied in this paper. This plot illustrates the fact that the ionizing clusters of \hii\/ regions located in the metal-rich inner regions of spiral galaxies are characterized by smaller effective temperatures than those further away from the galactic nuclei (\citealt{Bresolin:2002}). A metallicity dependence of the ionization parameter $U=Q_{H^0}/4\pi R_S^2nc$ (the ratio of ionizing photon density to atom density; $R_S$ is the Str\"omgren radius), as well as the relatively 
narrow range in the observed $U$ values, have been noted in studies of extragalactic \hii\/ regions several times (e.g.~\citealt{Dopita:1986,Bresolin:1999,Kewley:2002}), and have been confirmed by recent investigations of the integrated spectra of star-forming galaxies (\citealt{Nagao:2006,Maier:2006,Liang:2006}). The lower excitation of metal-rich nebulae 
can be explained by a combination of factors. First of all, stellar atmospheres of massive O stars become cooler with increasing metallicity, as a result of enhanced line and wind blanketing (\citealt{Massey:2005}).
Moreover, \citet{Dopita:2006b} have shown that the ionization parameter is determined by the ratio of the ionizing photons (a decreasing function of metallicity, due to enhanced stellar atmosphere opacities) to the mechanical energy input from the ionizing O stars (an increasing function of metallicity, 
since the photon momentum is more efficiently transferred to the stellar wind as the abundance of absorbers increases, 
\citealt{Bresolin:2004b}). The resulting metallicity dependence derived by Dopita et al., given approximately by $U\propto Z^{-0.8}$, is in good agreement with some empirical measurements. For example, \citet{Bresolin:1999} estimated a factor 4 decrease in $U$ varying the metallicity from 0.2\,$Z_\odot$ to 1.0\,$Z_\odot$.

\subsection{Temperature fluctuations}

The analysis based on the classic auroral line method provides
lower limits to the oxygen abundance in \hii\/ regions. Accounting for the presence of temperature fluctuations leads to  higher abundances relative to what is derived by assuming a constant temperature. Temperature fluctuations in nebulae can be estimated in different ways. As done here in the case of H1013, one can compare the temperature derived from the Balmer (or Paschen) continuum with $T$\oiii\/ (or other ionic temperature measured from an auroral-to-nebular line ratio). In virtually all of the Galactic and extragalactic \hii\/ regions with 12\,+\,(O/H)\,$\gtrsim$\,8.1
in which this comparison has been carried out, the former is found to be significantly lower than the latter (e.g.~\citealt{Peimbert:2000,Peimbert:2003,Garcia-Rojas:2006}). For a sample of metal-poor [\oh7.1-8.3] emission-line galaxies \citet{Guseva:2006} found, instead, that the two temperatures do not differ significantly, while \citet{Haegele:2006} found that only one out of three metal-poor [\oh7.9-8.0]
 \hii\/ galaxies shows significant temperature fluctuations.

It is also possible to estimate $t^2$ by requiring that the abundances obtained from collisionally excited lines and from recombination lines match (\citealt{Peimbert:1993}), and from the analysis of helium recombination lines (\citealt{Peimbert:2000,Peimbert:2005}). These methods have provided consistent results whenever a comparison has been attempted (\citealt{Esteban:2004,Garcia-Rojas:2004}). In extragalactic \hii\/ regions of high metal content the most feasible approach is the one adopted here for H1013, since metal recombination lines are difficult to detect. As the excitation of the \hii\/ regions decreases with increasing metallicity (Fig.~\ref{p}), the O\,{\sc ii} lines become fainter, despite the increase in the oxygen abundance. Even so, further efforts to detect metal recombination lines in high-metallicity nebulae, similar to what has been
done for \cii\lin4267 in H1013, would be very valuable. A somewhat higher spectral resolution ($R\sim5,000-10,000$) than the one used in this work on M101 will aid in the detection of these lines, as well as allow to disentangle lines belonging to a multiplet, such as O\,{\sc ii}.

The effects of temperature fluctuations on the derivation of nebular abundances need to be accounted for in the calibration of metallicity indicators based on strong lines. In Galactic and extragalactic \hii\/ regions an increase in oxygen abundance by a typical factor of 2 to 3 is found relative to 
the results obtained from collisionally excited lines under the assumption of $t^2\,=\,0$. Recently, \citet{Peimbert:2006} have compiled a list of \hii\/ regions where the abundances have been measured, mostly by the same authors, via metal recombination lines, as well as from collisionally excited lines, with the goal of establishing a preliminary calibration of $R_{23}$ as a function of (O/H) from the recombination line technique. Fig.~\ref{r23} shows the abundances measured in H1013 from this paper, for $t^2\,=\,0$ (large open square) and for $t^2\,=\,0.06$ (large full square), in the 12\,+\,log(O/H) vs.~$\log\,R_{23}$ plane. The \hii\/ regions in the \citet{Peimbert:2006} compilation are shown by the small open squares (abundances from collisionally excited lines, $t^2\,=\,0$) and by the small filled squares (abundances from metal recombinations lines or from an estimate of $t^2$ based on the Balmer continuum temperature). 
All objects plotted in the diagram belong to the upper branch of $R_{23}$, including those in the comparison sample of metal-rich \hii\/ regions from KBG03 and \citet[crosses, open triangles and open circles, respectively]{Bresolin:2004, Bresolin:2005}. The \hii\/ region H1013 in M101 is currently the extragalactic nebula that lies at the lowest $R_{23}$ value among those that currently have a $t^2$ estimate. Its oxygen and carbon abundances are among the highest measured so far in this relatively small sample (18 objects).

In Fig.~\ref{r23} the empirical measurements are compared with the $P$-method calibration of \citet{Pilyugin:2005} for two representative values of $P$ (0.2 and 0.6, solid curves). The nebular line-based abundances, derived by neglecting temperature fluctuations (open symbols), are well represented by these curves, and show the tendency mentioned above for a decreasing excitation parameter as the metallicity increases. The abundances derived from metal-recombination lines or accounting for temperature inhomogeneities (solid squares) are displaced upward by about 0.3 dex, approaching the
theoretical calculations from photoionization models, taken from \citet{Kobulnicky:2004} for two different values of the ionization parameter, $10^7$\,cm\,s$^{-1}$ and $10^8$\,cm\,s$^{-1}$   (these correspond to values of the dimensionless ionization parameter
$U\,=\,q/c$ of approximately $\log\,U=-3.5$ and $\log\,U=-2.5$, respectively).
Despite the difficulties that photoionization models have at reproducing emission lines of real nebulae that are very sensitive to \te, such as \oiii\lin4363, they are more successful with lines that have a smaller \te\/ dependence, such as \oii\lin3727 and \oiii\lin5007. This would explain the relatively good agreement found in Fig.~5 with the 
abundance determinations that account for temperature fluctuations (\citealt{Peimbert:2006}). Still, the problem remains that
standard photoionization models predict values for $t^2$ in \hii\/ regions that are very small, between 0.001 and 0.01, compared to observed values between 0.02 and 0.09 (\citealt{Peimbert:1995,Perez:1997}).

In future observations it will be important to extend the sample of objects where the abundances are obtained from recombination lines to lower values of $R_{23}$, down to $\log\,R_{23}\sim0$, if we want to establish a calibration of $R_{23}$ based on recombination lines and/or $t^2$ determinations that can be used at the highest metallicities. This is made somewhat problematic by the weakness of the metal recombination lines and by the difficulty in obtaining reliable Balmer continuum temperatures. 
Therefore, it is still valuable to measure nebular abundances from \oiii\lin4363 and other auroral lines in the high-metallicity regime, since, under the temperature fluctuation paradigm,  they appear to provide good estimates for the lower limits in abundance. 

\subsection{[N\,${\scriptstyle II}$]/[O\,${\scriptstyle II}$]}
\begin{figure}
\medskip
\plotone{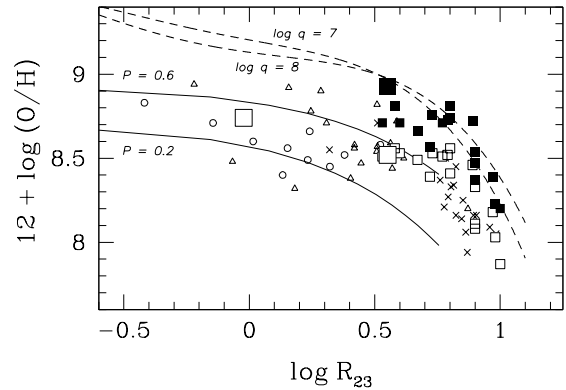}
\caption{The upper, high-metallicity branch of $R_{23}$. Samples of metal-rich \hii\/ regions are taken from KBG03 (crosses) and \citet[open triangles and open circles, respectively]{Bresolin:2004, Bresolin:2005}. Small squares represent Galactic and extragalactic \hii\/ regions, compiled by \citet{Peimbert:2006}, where abundances have been derived both from collisionally excited lines under the assumption $t^2\,=\,0$ (open symbols) and from metal recombination lines or an estimate of $t^2$ (full symbols). 
The \hii\/ regions H493 and H1013 studied in this paper are represented by the large square symbols.
The $P$-method calibration of \citet{Pilyugin:2005} is shown for $P\,=\,0.2$ and $P\,=\,0.6$ (full lines).
The $R_{23}$ calibration based on photoionization models by \citet{Kobulnicky:2004} is drawn for two values of the ionization parameter, $q=10^7$\,cm\,s$^{-1}$ ($\log\,U=-3.5$) and  $q=10^8$\,cm\,s$^{-1}$ ($\log\,U=-2.5$) (dashed lines). 
\label{r23}}
\end{figure}

\begin{figure}
\medskip
\plotone{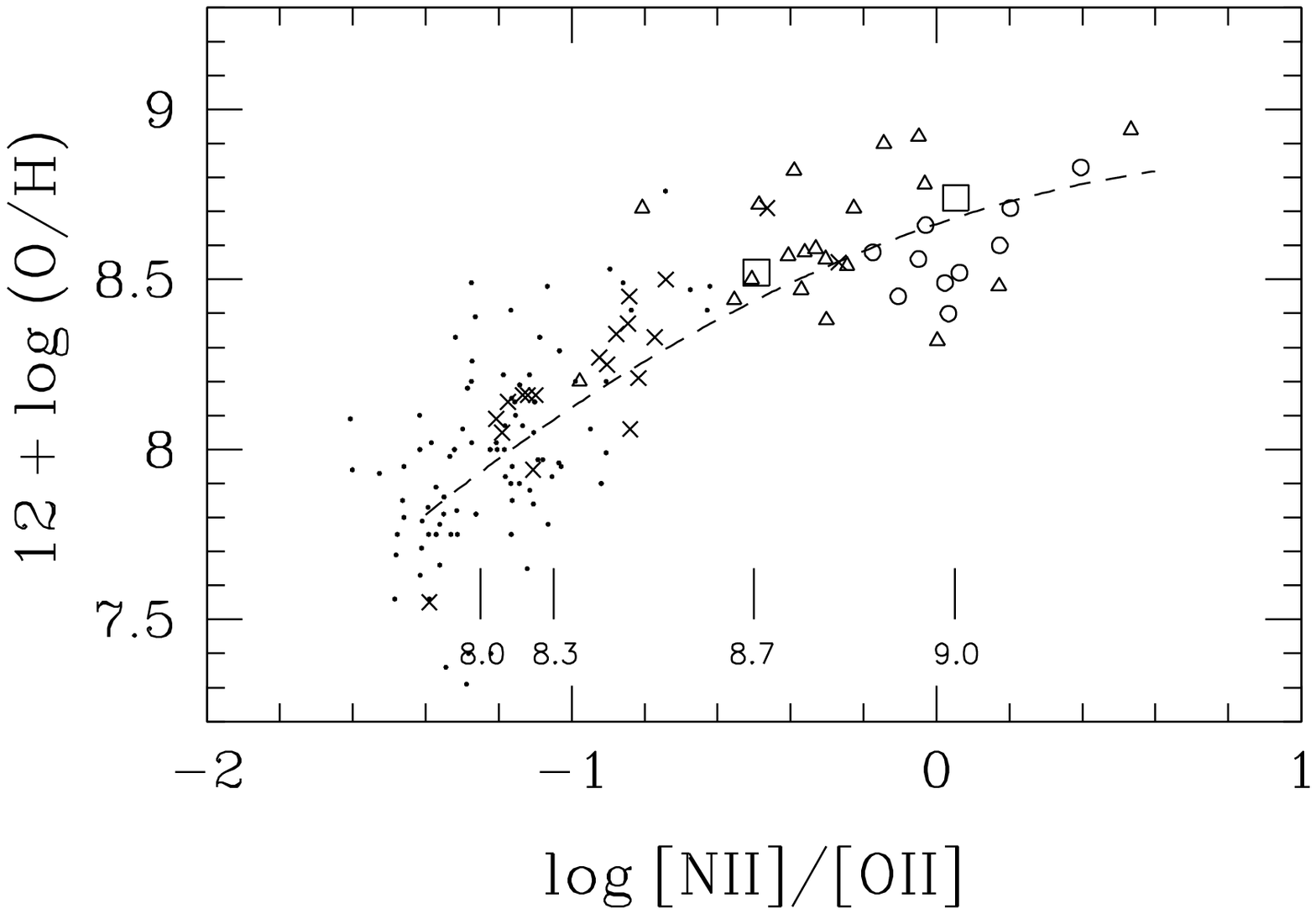}
\caption{Empirical relation between \nii\lin6583/\oii\lin3727 and oxygen abundance. The  \hii\/ regions plotted 
are the same as in Fig.~\ref{r23}, using the same symbols. The additional low metallicity points (small dots) are from \citet{Garnett:1997}, \citet{van-Zee:1998} and \citet{van-Zee:2006}. The simple fit to the data given by Eq.~9 is shown as a dashed curve. 
Four different values of O/H, taken from the models by \citet{Dopita:2006a}, are indicated at the bottom of the diagram.
\label{n2o2}}
\end{figure}

As mentioned before, the \nii/\oii\/ ratio can be used to remove the ambiguity between the lower and upper branches of $R_{23}$. In fact, \citet{Dopita:2000} showed that this ratio is an excellent abundance indicator by itself, superior to $R_{23}$ because of its monotonic dependence on metallicity and its insensitivity to the nebular excitation. In Fig.~\ref{n2o2} I show $x\,=\,\log($\nii\lin6583/\oii\lin3727) as a function of \te-based oxygen abundances for the same extragalactic \hii\/ regions included in Fig.~\ref{r23}, supplemented at the lower-abundance end by the samples studied by \citet{Garnett:1997}, \citet{van-Zee:1998} and \citet{van-Zee:2006} (\citealt{Perez-Montero:2005} showed a similar plot that includes a larger sample of objects). A simple fit to these data (dashed line) is:

\begin{equation}
\rm\oh 8.66 + 0.36\,x - 0.17\,x^2
\end{equation}

\noindent
According to the models by \citet[their Fig.~7]{Dopita:2006a}, a given $x$ corresponds with very good approximation to a unique value of the metallicity. The 12\,+\,log(O/H) values taken from the Dopita et al. models
are indicated in the bottom portion of Fig.~\ref{n2o2}. This helps to illustrate the typical result that abundance diagnostics calibrated via photoionization models provide oxygen abundances about 0.2 dex larger than those obtained from
direct methods. From what has been shown in Section~7.3, temperature fluctuations 
might offer an explanation for this discrepancy.
The empirical calibration of the \nii/\oii\/ abundance diagnostic also shows that, despite the fact that this line ratio is
not sensitive to the ionization parameter, the accuracy of the oxygen abundances that can be derived from it is not
better than what can be obtained from $R_{23}$. The $rms$ of the fit shown in Fig.~\ref{n2o2} is 0.20 dex, while limiting the fit to objects that are on the $R_{23}$ upper branch (the extragalactic \hii\/ regions shown in Fig.~\ref{r23}) gives $rms\sim0.15$, the same one  obtains by fitting these abundances as a function of $\log\,R_{23}$. This suggests that our 
poor knowledge of
additional parameters affecting the spectra of \hii\/ regions, such as the temperature and ionization structures,
currently limits the accuracy in determining abundances from strong-line diagnostics.

\bigskip

\section{Conclusions}

In this paper I have presented new deep spectra obtained with Keck/LRIS of four \hii\/ regions located in the central, metal-rich (O/H above solar) zone of the spiral galaxy M101. 
The main results obtained are summarized as following:\\[2mm]
\noindent
- electron temperatures have been measured from auroral-to-nebular line ratios for two \hii\/ regions, H493 and H1013. The classic analysis based on collisionally excited lines provides oxygen abundances of \oh8.74$\pm$0.17 and \oh8.52$\pm$0.05 for these two objects, respectively. These measurements extend the direct determination of the radial abundance gradient in M101 to about 3 kpc from the center ($R/R_0=0.10$).\\[1mm]
\noindent
- the C\,{\sc ii}\lin4267 recombination line has been measured in H1013. Accounting for the unseen C$^+$/H$^+$ fraction with the aid of photoionization models, a total carbon abundance \ch8.66 $\pm$ 0.07 (approximately 1.9 times the solar value) is found.\\[1mm]
\noindent
- for H1013 the comparison between the electron temperature derived from the Balmer jump, $T(Bac)\,=\,5000$~K, and the electron temperature derived from \oiii\lin4363/\lin5007, $T$\oiii\,=\,7700~K, leads to a mean square temperature fluctuation $t^2\,=\,0.06\pm 0.02$, with a volume averaged temperature $T_0\,=\,5500\pm 700$~K.\\[1mm]
\noindent
- correcting the abundances derived from the collisionally excited lines for the effect of temperature inhomogeneities 
increases the oxygen abundance in H1013 by almost 0.4 dex, to \oh8.93 $\pm$ 0.05. Combining this result with the carbon abundance derived from C\,{\sc ii}\lin4267 gives log(C/O)\, =\, $-0.27\pm0.11$.\\[1mm]
\noindent
- these results indicate that there is still no empirical verification of the strong abundance biases from temperature gradients predicted theoretically for metal-rich \hii\/ regions.\\[1mm]
\noindent
- temperature fluctuations can quantitatively explain the current abundance discrepancy of $\sim$0.2-0.3 dex between empirical indicators and theoretical models.\\[2mm]
\noindent
Future observational work needs to target high-metallicity, low excitation \hii\/ regions in the local Universe with medium resolution spectroscopy, in search for metal recombination lines. From these, chemical abundances that are virtually 
independent of the temperature structure of \hii\/ regions can be derived. This would allow more accurate calibrations of strong-line metallicity indicators.


\acknowledgments

It is a pleasure to thank Manuel Peimbert for encouragement and suggestions.

\bibliography{M101inner}





\end{document}